\documentclass[12pt]{article}
\usepackage{makeidx}
\usepackage{amsmath}
\usepackage{amsfonts}
\usepackage{amssymb}
\usepackage{epsfig}

\newcounter{defin}
\newcounter{lemma}
\newcounter{theorem}
\newcounter{corollary}
\newcounter{proposition}
\newcounter{example}
\newenvironment{lemma}{\par\refstepcounter{lemma}     \textbf{Lemma \thelemma.} }{\rm\par}

\begin{document}

\title{On estimates of trace-norm distance between quantum Gaussian states}
\author{Holevo~A.\,S. \\
Steklov Mathematical Institute, \\
Russian Academy of Sciences, Moscow, Russia}
\date{}
\maketitle

\begin{abstract}
In the paper \cite{lami} estimates for the trace-norm distance between two
quantum Gaussian states of CCR in terms of the mean vectors and covariance matrices
were derived and used to evaluate the number of elements in the $\varepsilon
-$net in the set of energy-constrained Gaussian states. In the present paper
we obtain different estimates based on a fidelity-like
quantity which we call \textit{states overlap} \cite{tmf}. Our proof is more direct
leading to estimates which are sometimes even more stringent, especially in
the cases of pure or gauge-invariant states. They do not depend on number of
modes and hence can be extended to the case of bosonic field with infinite
number of modes. These derivations are not aimed to replace the useful
inequalities from \cite{lami}; they just show an alternative approach to the
problem leading to different results.  In the Appendix we briefly recall our results from
 \cite{tmf2} concerning estimates of the states overlap for general fermionic Gaussian states
of CAR.

The problems studied in this paper can
be considered as a noncommutative analog of estimation of the total variance
distance between Gaussian probability distributions in the classical
probability theory.

\end{abstract}

\section{Introduction and results}

In this paper we obtain estimates for trace-norm distance $\left\Vert \rho
_{1}-\rho _{2}\right\Vert _{1}$ between two quantum Gaussian states in terms
of their parameters -- the mean vectors and covariance matrices. To our
knowledge, such kind of estimates were first derived in \cite{lami}, where
they were used to evaluate the number of elements in the $\varepsilon -$net
in the set of energy-constrained Gaussian states implying that such a state
can be efficiently learned via quantum tomography. It was also noticed that
such estimates are of an independent interest. Indeed, as an example, they
could be useful in investigations concerning characterization of quantum
Gaussian observables \cite{tvp}. The problem studied in the present paper
can be also considered as a noncommutative analog of estimation of the total
variance distance between Gaussian probability distributions in the
classical probability theory \cite{dan}, \cite{108}, \cite{109}.

Concerning the method of the proof, the authors of \cite{lami} observe:
``Initially, one might believe that proving our theorem 3 would be
straightforward by bounding the trace distance using the fidelity and
leveraging the established formula for the fidelity between Gaussian states
\cite{pir}. However, this approach turns out to be highly non-trivial due to
the complexity of such fidelity formula \cite{pir}, which makes it
challenging to derive a bound based on the norm distance between the first
moments and the covariance matrices. Instead, our proof technique directly
addresses the trace distance without relying on fidelity. It involves a
meticulous analysis based on properties of Gaussian channels and recently
demonstrated properties of the energy-constrained diamond norm...''

A hypothetical proof based on the fidelity $\mathrm{Tr}\left\vert \sqrt{\rho
_{1}}\sqrt{\rho _{2}}\right\vert $ indeed looks to be too difficult. In the
present paper, instead of the fidelity we try a simpler quantity $\mathrm{Tr}%
\sqrt{\rho _{1}}\sqrt{\rho _{2}}$, which we call here quantum \textit{states
overlap}\footnote{Some authors call the overlap a different quantity $\mathrm{Tr}%
\rho _{1}\rho _{2}$.}, and use the corresponding estimate of the trace-norm distance from
our paper \cite{tmf}. This allows us for a more straightforward proof of the
estimates which are sometimes even more stringent, especially in the cases
of pure states. They do not depend on number of modes and
hence can be extended to the case of bosonic field with infinite number of
modes. However these derivations are not aimed to replace the useful
inequalities from \cite{lami} which are adjusted for the purposes of that
important paper; they just show a possibility of an alternative approach to
the problem leading to different estimates\footnote{After the first posting of the present paper,
the sequel \cite{lami2} of  \cite{lami}  appeared where the estimates of  \cite{lami} were substantially improved
by using still a different gradient method. Alternative estimates based on quantum Pinsker inequality targeted to
Hamiltonian learning were demonstrated in \cite{rouze}.}.

Throughout this paper $\Vert \cdot \Vert _{1}$ denotes \textit{the trace norm%
} of a matrix or an operator on a Hilbert space, $\Vert \cdot \Vert _{2}$ --
either \textit{the Hilbert-Schmidt norm} of an operator or \textit{Euclidean
norm} of a real vector, and $\Vert \cdot \Vert $ --\textit{\ the operator
norm.}

We consider two quantum Gaussian states $\rho _{m_{1},\alpha }$ and $\rho
_{m_{2},\beta }$ with the mean vectors $m_{1},m_{2}$ and the covariance
matrices $\alpha $, $\beta $ (see sec. \ref{sec:ccr}). We introduce the
numbers $a=\Vert \alpha \Vert $ and $b=\Vert \beta \Vert $. Then our
estimates are: for pure Gaussian states
\begin{equation}
\left\Vert \rho _{m_{1},\alpha }-\rho _{_{m_{2},\beta }}\right\Vert _{1}\leq
\sqrt{2 \left( a+b\right) \left\Vert m_{1}-m_{2}\right\Vert _{2}^{2}+\left\Vert
\alpha -\beta \right\Vert _{2}^{2}},  \label{E1}
\end{equation}%
and for gauge invariant (\textquotedblleft passive\textquotedblright )
states ($m_{1}=m_{2}=0$, $\alpha $ and $\beta $ commute with the commutation
matrix $\Delta $)
\begin{equation}
\left\Vert \rho _{\alpha }-\rho _{_{\beta }}\right\Vert _{1}\leq \sqrt{2[
\left( a+b\right) \left\Vert \alpha -\beta \right\Vert _{1}-\left\Vert
\alpha -\beta \right\Vert _{2}^{2}]}.  \label{E2}
\end{equation}%
For arbitrary Gaussian state we were able to derive the following estimate
\begin{equation}
\left\Vert \rho _{m_{1},\alpha }-\rho _{_{m_{2},\beta }}\right\Vert _{1}\leq
\label{E3}
\end{equation}%
\begin{equation*}
2\sqrt{\left( a+b\right) \left\Vert m_{1}-m_{2}\right\Vert _{2}^{2}+%
\frac{1}{2}\left\Vert \alpha -\beta \right\Vert _{2}^{2}+\min \left\{
P(a,b),P(b,a)\right\} \left\Vert \alpha -\beta \right\Vert _{1}},
\end{equation*}%
where $P$ is a polynomial with coefficients explicitly given in the proof,
see sec. \ref{sec:proofs}. Comparing with the formulas for pure and
gauge-covariant states, we conjecture that there might be a more clever way
to use the state overlap in the case of general Gaussian states to improve
the inequality (\ref{E3}).

The proofs are presented in sec. \ref{sec:proofs}. In the concluding sec. %
\ref{sec:disc} we compare our estimates with those obtained in \cite%
{lami}.  In the Appendix we briefly recall our results from
 \cite{tmf2} concerning estimates of state overlap for general fermionic Gaussian states
of CAR.

\section{The algebra of CCR}

\label{sec:ccr}

We first recall some standard material from the books \cite{asp}, \cite{h}.
Let $(Z,\Delta )$ be a finite-dimensional symplectic vector space with $Z=%
\mathbb{R}^{2s}$ and the canonical commutation matrix
\begin{equation}
\Delta =\mathrm{diag}\left[
\begin{array}{cc}
0 & 1 \\
-1 & 0%
\end{array}%
\right] _{j=1,\dots ,s},  \label{delta}
\end{equation}%
defining the symplectic form $\Delta (z,z^{\prime })=z^{t}\Delta z^{\prime }.
$ In what follows the Hilbert space $\mathcal{H}$ is the space of an
irreducible representation $z\rightarrow W(z);\,z\in Z,$ of the Weyl
canonical commutation relations (CCR)
\begin{equation}
W(z)W(z^{\prime })=\exp [-\frac{i}{2}\Delta (z,z^{\prime })]\,W(z+z^{\prime
}),  \label{eq17}
\end{equation}%
where $W(z)=\exp i\,Rz$ are the unitary operators with the generators
\begin{equation}
Rz=\sum_{j=1}^{s}(x_{j}q_{j}+y_{j}p_{j}),\quad z=[x_{j}\,\,y_{j}]_{j=1,\dots
,s}^{t}
\end{equation}%
and $R=\left[ q_{1},p_{1},\dots ,q_{s},p_{s}\right] $ is the row vector of
the bosonic position-momentum observables, satisfying the Heisenberg
commutation relations on an appropriate domain. 

The quantum Fourier transform of a trace class operator $\rho $ is defined
as $f(z)=\mathop{\rm Tr}\nolimits\rho\, W(z)$.
The quantum Parceval formula holds \cite{asp}:%
\begin{equation}
\mathop{\rm Tr}\nolimits\rho ^{\ast }\sigma =\int \,\overline{\mathop{\rm Tr}%
\nolimits\rho W(z)}\mathop{\rm Tr}\nolimits\sigma W(z)\frac{d^{2s}z}{\left(
2\pi \right) ^{s}} .  \label{parc}
\end{equation}

The covariance matrix of a state with finite second moments is defined as
\begin{equation*}
\alpha =\mathrm{Re}\,\mathrm{Tr}\left( R-m\right) ^{t}\rho \left( R-m\right)
,
\end{equation*}%
where $m=\mathrm{Tr}\rho R$ is the row-vector of the first moments (the mean
vector). Matrix $\alpha $ is a real symmetric $2s\times 2s $-matrix
satisfying
\begin{equation}
\alpha \geq \pm \frac{i}{2}\Delta .  \label{ur}
\end{equation}%

The condition (\ref{ur}) implies that matrix $\alpha $ is positive definite.
There is a theorem in linear algebra saying that a positive definite matrix
can be represented as
\begin{equation*}
\alpha =\,S^{t}\mathrm{diag}\left[
\begin{array}{cc}
a_{j} & 0 \\
0 & a_{j}%
\end{array}%
\right] _{j==1,\dots ,s} S,
\end{equation*}%
where $S$ is a symplectic matrix ($S^{t}\Delta S=\Delta $) and $a_{j}>0$
(see e.g. \cite{asp}). For a positive definite matrix $\alpha $ the
inequality
\begin{equation}
\alpha +\frac{1}{4}\Delta \alpha ^{-1}\Delta \geq 0.  \label{ur2}
\end{equation}%
is equivalent to (\ref{ur}), because both (\ref{ur}) and (\ref{ur2}) express
the fact that symplectic eigenvalues of matrix $\alpha $ satisfy $a_{j}\geq
1/2$, $j=1,\dots ,s.$

A \textit{Gaussian state} $\rho _{m,\alpha }$ is determined by its quantum
characteristic function%
\begin{equation}
\mathrm{Tr}\,\rho _{m,\alpha }W(z)=\exp \left( im^{t}z-\frac{1}{2}%
z^{t}\alpha z\right) .  \label{gs}
\end{equation}%
Here $\alpha $ is the covariance matrix and $m$ is the mean vector of the
state. For a centered state $(m=0)$ we denote $\rho _{\alpha }=\rho
_{0,\alpha }.$

The equality
\begin{equation}
\alpha +\frac{1}{4}\Delta \alpha ^{-1}\Delta =0  \label{pure}
\end{equation}%
in (\ref{ur2})\ holds if and only if $\rho _{m,\alpha }$ is a pure Gaussian
state $(a_{j}=1/2; j=1,\dots ,s)$.

Let $\alpha $ be a covariance matrix of a Gaussian state, that is a real
symmetric matrix satisfying the inequality (\ref{ur}). It is thus positive
definite, hence defines the inner product $\alpha (z,z^{\prime
})=z^{t}\alpha z^{\prime }$ in $Z$ making it the real Hilbert space $%
(Z,\alpha ).$ If $T$ is an operator in $(Z,\alpha )$, then its adjoint is $%
\left( T\right) _{\alpha }^{\ast }=\alpha ^{-1}T^{t}\alpha .$ We call $T$ $%
\alpha -$symmetric ($\alpha -$skew-symmetric) if $T=\left( T\right) _{\alpha
}^{\ast }$ ($T=-\left( T\right) _{\alpha }^{\ast }$), which amounts to $%
\alpha T=T^{t}\alpha $ (resp. $\alpha T=-T^{t}\alpha $). An $\alpha -$%
symmetric operator $T$ is $\alpha -$positive if $\ \alpha (z,Tz)\geq 0$ for
all $z,$ which amounts to $\alpha T\geq 0.$ For such an operator the square
root $\sqrt{T}$ is the unique $\alpha -$positive operator satisfying $\left(
\sqrt{T}\right) ^{2}=T.$

Consider the operator\footnote{%
In the case (\ref{delta}) we have $\Delta ^{-1}=-\Delta$, but we retain the
notation $\Delta ^{-1}$ having in mind more general situation where $\Delta$
is an arbitrary nondegenerate skew-symmetrc matrix. Our derivations then
remain valid with appropriate modifications.} $A=\Delta ^{-1}\alpha $ which
represents the bilinear form $\alpha (z,z^{\prime })=z^{t}\alpha z^{\prime }$
in the symplectic space $(Z,\Delta ):$%
\begin{equation*}
\alpha (z,z^{\prime })=\Delta (z,Az^{\prime }),\quad z,z^{\prime }\in Z.
\end{equation*}%
This operator is skew-symmetric in the Euclidean space $(Z,\alpha )$, hence $%
I_{2s}+\left( 2\Delta ^{-1}\alpha \right) ^{-2}$ is $\alpha -$symmetric.
Moreover it is $\alpha -$positive as
\begin{equation}
\alpha \left[ I_{2s}+\left( 2\Delta ^{-1}\alpha \right) ^{-2}\right] =\alpha
+\frac{1}{4}\Delta \alpha ^{-1}\Delta \geq 0,  \label{ineq3}
\end{equation}%
due to (\ref{ur2})$.$ Hence the square root%
\begin{equation}
\Upsilon _{\alpha }=\sqrt{I_{2s}+\left( 2\Delta ^{-1}\alpha \right) ^{-2}}
\label{ups}
\end{equation}%
is defined and satisfies%
\begin{equation}
\alpha \Upsilon _{\alpha }=\Upsilon _{\alpha }^{t}\alpha \geq 0.  \label{per}
\end{equation}

Later we will need the following important matrix
\begin{equation}
\hat{\alpha}\equiv \alpha \left( I_{2s}+\Upsilon _{\alpha }\right) =\alpha
\left( I_{2s}+\sqrt{I_{2s}+\left( 2\Delta ^{-1}\alpha \right) ^{-2}}\right) .
\label{ahat}
\end{equation}%
The matrix $\hat{\alpha}$ is\textit{\ }$\alpha -$symmetric and satisfies
\begin{equation}
-\frac{1}{4}\Delta \hat{\alpha}^{-1}\Delta =2\alpha -\hat{\alpha}.
\label{aminua}
\end{equation}%
Indeed by using (\ref{ahat}) and noticing that $\left( I_{2s}+\Upsilon
_{\alpha }\right) \left( I_{2s}-\Upsilon _{\alpha }\right) =-\left( 2\Delta
^{-1}\alpha \right) ^{-2}$ we have%
\begin{eqnarray}
\hat{\alpha}^{-1} &=&-4\Delta ^{-1}\alpha \left( I_{2s}-\Upsilon _{\alpha
}\right) \Delta ^{-1}  \label{hat1} \\
&=&-4\Delta ^{-1}\alpha \Delta ^{-1}+4\Delta ^{-1}\alpha \Upsilon _{\alpha
}\Delta ^{-1}  \label{hat2} \\
&=&-4\Delta ^{-1}\left( 2\alpha -\hat{\alpha}\right) \Delta ^{-1},
\label{hat3}
\end{eqnarray}%
where we also used the fact that $\Upsilon _{\alpha }$ is a function of $%
\Delta ^{-1}\alpha $ and hence commutes with it. But this is equivalent to (%
\ref{aminua}).

In particular, (\ref{per}) and (\ref{aminua}) imply
\begin{equation}
\alpha \leq \hat{\alpha}\leq 2\alpha ,  \label{ineq5}
\end{equation}%
because
\begin{equation}
0\leq \frac{1}{4}\Delta ^{t}\hat{\alpha}^{-1}\Delta =-\frac{1}{4}\Delta \hat{%
\alpha}^{-1}\Delta .  \label{unc2}
\end{equation}%
Also from (\ref{hat1})%
\begin{equation*}
\alpha \left( I_{2s}-\Upsilon _{\alpha }\right) =-\frac{1}{4}\Delta \hat{%
\alpha}^{-1}\Delta \geq 0
\end{equation*}%
hence%
\begin{equation}
\alpha \Upsilon _{\alpha }\leq \alpha .  \label{aga}
\end{equation}

Introducing the number $a=\left\Vert \alpha \right\Vert ,$ we therefore have
the estimates
\begin{equation}
a=\left\Vert \alpha \right\Vert \leq \left\Vert \hat{\alpha}\right\Vert \leq
2\left\Vert \alpha \right\Vert =2a.  \label{norm0}
\end{equation}%
From (\ref{ur2}) $\alpha ^{-1}\leq -4\Delta \alpha \Delta =4\Delta
^{t}\alpha \Delta $ hence taking into account that $\left\Vert \Delta
\right\Vert =1,$
\begin{equation}
a^{-1}=\left\Vert \alpha \right\Vert ^{-1}\leq \left\Vert \alpha
^{-1}\right\Vert \leq 4\left\Vert \alpha \right\Vert =4a.  \label{norm1}
\end{equation}%
By (\ref{ineq5}), $\frac{1}{2}\alpha ^{-1}\leq \hat{\alpha}^{-1}\leq \alpha
^{-1},$ hence
\begin{equation}
\frac{1}{2a}\leq \frac{1}{2}\left\Vert \alpha ^{-1}\right\Vert \leq
\left\Vert \hat{\alpha}^{-1}\right\Vert \leq \left\Vert \alpha
^{-1}\right\Vert \leq 4a.  \label{norm2}
\end{equation}%
\bigskip

\section{\protect\bigskip Proofs}

\label{sec:proofs}

Our approach uses the second inequality in%
\begin{equation}
2\left( 1-\mathrm{Tr}\sqrt{\rho _{1}}\sqrt{\rho _{2}}\right) \leq \left\Vert
\rho _{1}-\rho _{2}\right\Vert _{1}\leq 2\sqrt{1-\left( \mathrm{Tr}\sqrt{%
\rho _{1}}\sqrt{\rho _{2}}\right) ^{2}},  \label{basic}
\end{equation}%
where $\rho _{1},\rho _{2}$ are density operators.  The proof
given in \cite{tmf} is partly based on estimates from \cite{powers}.

The unitary displacement operators $D(z)=W(-\Delta ^{-1}z)$ satisfy the
equation that follows from the canonical commutation relations (\ref{eq17})
\begin{equation}
D(z)^{\ast }W(w)D(z)=\exp \left( iw^{t}z\right) W(w),  \label{dis}
\end{equation}%
implying
\begin{equation*}
\rho _{m_{1},\alpha }=D(m_{1})\rho _{\alpha }D(m_{1})^{\ast },\quad \rho
_{m_{2},\beta }=D(m_{1})\rho _{_{m,\beta }}D(m_{1})^{\ast },
\end{equation*}%
where $m=m_{2}-m_{1}.$ Hence
\begin{equation*}
\left\Vert \rho _{m_{1},\alpha }-\rho _{m_{2},\beta }\right\Vert
_{1}=\left\Vert \rho _{\alpha }-\rho _{_{m,\beta }}\right\Vert _{1},
\end{equation*}%
and in what folows we will deal with the second quantity.

From \cite{tmf} we know that the quantum Fourier transform of $\sqrt{\rho
_{m,\alpha }}$ is
\begin{equation}
\mathrm{Tr}\,\sqrt{\rho _{m,\alpha }}W(z)=\sqrt[4]{\det \left( 2\hat{\alpha}%
\right) }\exp \left( im^{t}z-\frac{1}{2}z^{t}\hat{\alpha}z\right) ,
\label{sqr}
\end{equation}%
where $\hat{\alpha}$ is given by (\ref{ahat}).

Applying the Parceval relation (\ref{parc}) we arrive at the formula
\begin{equation}  \label{overl}
\mathrm{Tr}\sqrt{\rho _{\alpha }}\sqrt{\rho _{m,\beta }}=\frac{\left( \det
\hat{\alpha}\det \hat{\beta}\right) ^{1/4}}{\det \sigma (\hat{\alpha},\hat{%
\beta})^{1/2}}\exp \left( -\frac{1}{4}m^{t}\sigma (\hat{\alpha},\hat{\beta}%
)^{-1}m\right) ,
\end{equation}%
\begin{equation*}
\sigma (\hat{\alpha},\hat{\beta})=\frac{\hat{\alpha}+\hat{\beta}}{2}.
\end{equation*}

Let us evaluate the factor before the exponent:
\begin{eqnarray*}
&&\frac{\left( \det \hat{\alpha}\det \hat{\beta}\right) ^{1/4}}{\det \sigma (%
\hat{\alpha},\hat{\beta})^{1/2}} \\
&=&\left[ \det \left( \frac{1+\hat{\alpha}^{-1}\hat{\beta}}{2}\right) \det
\left( \frac{1+\hat{\beta}^{-1}\hat{\alpha}}{2}\right) \right] ^{-1/4} \\
&=&\left[ \det \left( \frac{T+2+T^{-1}}{4}\right) \right] ^{-1/4},
\end{eqnarray*}%
where $T=\hat{\alpha}^{-1}\hat{\beta}$ is positive operator in the Hilbert
space $Z_{\hat{\alpha}}.$ Also
\begin{equation*}
d(\alpha ,\beta )=\frac{1}{4}\left( T-2+T^{-1}\right)
\end{equation*}
is positive operator in $Z_{\hat{\alpha}},$ therefore
\begin{eqnarray*}
&&\left[ \det \left( \frac{T+2+T^{-1}}{4}\right) \right] ^{-1/4} \\
&=&\exp \left[ -\frac{1}{4}\mathrm{Tr}\ln \left( \frac{T+2+T^{-1}}{4}\right) %
\right] \\
&=&\exp \left[ -\frac{1}{4}\mathrm{Tr}\ln \left( I+d(\alpha ,\beta )\right) %
\right].
\end{eqnarray*}%
By using the inequalities $\frac{\ln (1+c)}{c}x\leq \ln \left( 1+x\right)
\leq x$ for $0\leq x\leq c,$ we obtain%
\begin{eqnarray}  \label{first}
&& \exp \left[ -\frac{1}{4}\mathrm{Tr\,}d(\alpha ,\beta )-\frac{1}{2}m^{t}\left( \hat{%
\alpha}+\hat{\beta}\right) ^{-1}m\right] \leq \mathrm{Tr}\sqrt{\rho _{\alpha
}}\sqrt{\rho _{m,\beta }} \\
&\leq &\exp \left[ -\frac{1}{4}\frac{\ln (1+\left\Vert d(\alpha ,\beta
)\right\Vert )}{\left\Vert d(\alpha ,\beta )\right\Vert }\mathrm{Tr\,}%
d(\alpha ,\beta )-\frac{1}{2}m^{t}\left( \hat{\alpha}+\hat{\beta}\right) ^{-1}m\right] .
\label{second}
\end{eqnarray}

The first inequality implies%
\begin{equation}\label{ineq6}
1-\left( \mathrm{Tr}\sqrt{\rho _{\alpha }}\sqrt{\rho _{m,\beta }}\right)
^{2} \leq m^{t}\left( \hat{\alpha}+\hat{\beta}\right) ^{-1}m+\frac{1}{2}%
\mathrm{Tr\,}d(\alpha ,\beta ).
\end{equation}%
But%
\begin{equation*}
\mathrm{Tr\,}d(\alpha ,\beta )=\frac{1}{4}\mathrm{Tr}\left( \hat{\alpha}-%
\hat{\beta}\right) \left( \hat{\beta}^{-1}-\hat{\alpha}^{-1}\right) .
\end{equation*}%
The second inequality in (\ref{basic}) then implies%
\begin{equation}
\left\Vert \rho _{\alpha }-\rho _{_{m,\beta }}\right\Vert _{1}\leq 2\sqrt{%
m^{t}\left( \hat{\alpha}+\hat{\beta}\right) ^{-1}m+\frac{1}{2}\mathrm{Tr\,}%
d(\alpha ,\beta )}.  \label{basic2}
\end{equation}

Let us evaluate the expression $\mathrm{Tr}\left( \hat{\alpha}-\hat{\beta}%
\right) \left( \hat{\beta}^{-1}-\hat{\alpha}^{-1}\right) .$ From (\ref{hat2}%
)
\begin{equation*}
\hat{\alpha}^{-1} = -4\Delta ^{-1}\alpha \Delta ^{-1}+4\Delta ^{-1}\alpha
\Upsilon _{\alpha }\Delta ^{-1}.
\end{equation*}%
By using this and (\ref{ahat}) we have
\begin{eqnarray}
&&\mathrm{Tr\,}d(\alpha ,\beta )=\frac{1}{4}\mathrm{Tr}\left( \hat{\alpha}-\hat{\beta}\right) \left( \hat{\beta}^{-1}-%
\hat{\alpha}^{-1}\right)  \label{ea} \\
&=&\mathrm{Tr}[\left( \alpha -\beta \right) +\left( \alpha \Upsilon
_{\alpha }-\beta \Upsilon _{\beta }\right)] \Delta ^{-1} [\left( \alpha
-\beta \right) - \left( \alpha \Upsilon _{\alpha }-\beta \Upsilon _{\beta
}\right)]\Delta ^{-1}  \notag \\
&=&\mathrm{Tr}[\left( \alpha -\beta \right) \Delta ^{-1}]^{2}-\mathrm{Tr}%
[\left( \alpha \Upsilon _{\alpha }-\beta \Upsilon _{\beta }\right) \Delta
^{-1}]^{2}  \label{eb} \\
&\leq &\mathrm{Tr}[\alpha -\beta ]^{2}+\mathrm{Tr}[\alpha \Upsilon
_{\alpha }-\beta \Upsilon _{\beta }]^{2},  \label{ec}
\end{eqnarray}%
where in (\ref{ec}) we used the Cauchy-Schwarz inequality for trace and the
fact that $\left\Vert \Delta ^{-1}\right\Vert =1.$

\subsection{Pure states}

In the case of general pure states there is exact expression (see e.g. Lemma
10.9 in \cite{h}) which can be rewritten in the form%
\begin{equation}
\left\Vert \rho _{1}-\rho _{2}\right\Vert _{1} = 2\sqrt{1-\mathrm{Tr}\rho
_{1}\rho _{2}}=2\sqrt{1-\mathrm{Tr}\sqrt{\rho _{1}}\sqrt{\rho _{2}}}..
\label{basic3}
\end{equation}%
Also for Gaussian pure states the relation (\ref{pure}) implies $\left(
2\Delta ^{-1}\alpha \right) ^{2}=-I_{2s},$ that is $\Upsilon _{\alpha }=0$
and $\hat{\alpha}=\alpha .$ Similarly $\hat{\beta}=\beta $ and $\Upsilon
_{\beta }=0$. Then (\ref{ec}) with (\ref{first}) imply%
\begin{equation*}
\left\Vert \rho _{\alpha }-\rho _{_{m,\beta }}\right\Vert _{1}\leq \sqrt{%
m^{t}\left(\frac{\alpha +\beta}{2} \right) ^{-1}m+\mathrm{Tr}\left( \alpha
-\beta \right) ^{2}}.
\end{equation*}%
Since $\sigma = \frac{\alpha +\beta}{2}$ satisfies (\ref{ur}) along with $\alpha, \beta$, it
satisfies also (\ref{norm1}), hence
\begin{equation}
m^{t}\sigma ^{-1}m \leq  \left\Vert
\sigma^{-1}\right\Vert
\left\Vert m\right\Vert _{2}^{2}
\leq 4 \|\sigma\| \left\Vert m\right\Vert _{2}^{2} \leq 2(a+b)\left\Vert m\right\Vert _{2}^{2},
\label{mean}
\end{equation}%
where $a=\left\Vert \alpha \right\Vert ,b=\left\Vert \beta \right\Vert .$
Thus for pure Gaussian states we obtain%
\begin{equation}
\left\Vert \rho _{\alpha }-\rho _{_{m,\beta }}\right\Vert _{1}\leq \sqrt{%
2 \left( a+b\right) \left\Vert m\right\Vert _{2}^{2}+\left\Vert \alpha
-\beta \right\Vert _{2}^{2}},  \label{pureest}
\end{equation}%
that is our first estimate (\ref{E1}).


\subsection{Gauge-invariant states}

Operator $J$ in $(Z,\Delta )$ is called operator of complex structure if $%
J^{2}=-I_{2s},$ where $I_{2s}$ is the identity operator in $Z$, and it is $%
\Delta -$positive in the sense that
\begin{equation}
\Delta J=-J^{t}\Delta ,\quad \Delta J\geq 0.  \label{comstr}
\end{equation}%
A distinguished complex structure is given by $J_{\Delta }=\Delta ^{-1}.$
For any complex structure $J$ we have $J=S^{t}J_{\Delta }S,$ where $S$ is an
appropriate symplectic transformation.

The Gaussian state is gauge-invariant with respect to $J$ if the mean vector
$m=0$ and the operator $A=\Delta ^{-1}\alpha $ satisfies $[A,J]=0$ which is
equivalent to $\alpha J=-J^{t}\alpha .$ In the case $J=J_{\Delta }=\Delta
^{-1}$ this amounts to $[\alpha ,\Delta ]=0.$

Since all the complex structures are symplectically equivalent to $J_{\Delta
}=\Delta ^{-1},$ we can reduce to the case $J=J_{\Delta }$. Let us apply the
expression (\ref{eb}) in this case. Then $\Upsilon _{\alpha }=\sqrt{%
I_{2s}-\left( 2\alpha \right) ^{-2}}$ and $\hat{\alpha}=\alpha +\sqrt{\alpha
^{2}-1/4}$ and similarly for $\beta .$ Taking into account that $J_{\Delta }$
commutes with $\alpha $ and $\beta $ and $J_{\Delta }^{2}=-I_{2s}$, the
expression (\ref{eb}) turns into%
\begin{equation}
\mathrm{Tr\,}d(\alpha ,\beta )=-\mathrm{Tr}[\alpha -\beta ]^{2}+\mathrm{Tr}[\sqrt{\alpha ^{2}-1/4}-\sqrt{%
\beta ^{2}-1/4}]^{2}.  \label{sqrt}
\end{equation}%
For the second term, the operators $T_{1}=\sqrt{\alpha ^{2}-1/4},T_{2}=\sqrt{%
\beta ^{2}-1/4}$ are positive in the real Hilbert space $(Z, I)$ equipped
with inner product $(z,z^{\prime })= z^{t}z^{\prime }$ (corresponding to $%
\alpha =I_{2s}$). Hence we can use the inequality (1.2) from \cite{tmf} (see
also \cite{powers})
\begin{equation}  \label{21}
\left\Vert T_{1}-T_{2}\right\Vert _{2}^{2}\leq \left\Vert
T_{1}^{2}-T_{2}^{2}\right\Vert _{1}
\end{equation}%
implying that the second term in r.h.s. of (\ref{sqrt}) is upperbounded by%
\begin{eqnarray}
\left\Vert \alpha ^{2}-\beta ^{2}\right\Vert _{1} &=&\left\Vert \left(
\alpha -\beta \right) \circ \left( \alpha +\beta \right) \right\Vert _{1}\notag \\
&\leq &\left( \left\Vert \alpha \right\Vert +\left\Vert \beta \right\Vert
\right) \left\Vert \alpha -\beta \right\Vert _{1}\label{uper} \\
&=&\left( a+b\right) \left\Vert \alpha -\beta \right\Vert _{1},\notag
\end{eqnarray}%
where $\circ$ denoted the Jordan product. Inserting in (\ref{sqrt}) and using (\ref{basic2}) we
obtain the final estimate (\ref{E2}) for the trace-norm distance of
gauge-invariant states. %

\subsection{``Brute force'' estimate for arbitrary Gaussian states}

Consider the real Hilbert space $(Z,\alpha ).$ We wish to compare Schatten
norms of operators on $(Z,\alpha )$ (called $\alpha -$norms) with norms of
the same operators in the Hilbert space $(Z,I)$ equipped with the inner
product $z^{t}z^{\prime }$. Recall that the adjoint in $(Z,\alpha )$ is $%
\left( T\right) _{\alpha }^{\ast }=\alpha ^{-1}T^{t}\alpha ,$ hence the
modulus of the operator is $\left\vert T\right\vert _{\alpha }=\sqrt{\alpha
^{-1}T^{t}\alpha T}.$ Therefore Schatten $\alpha -$norms for $p\geq 1$ are%
\begin{equation*}
\left\Vert T\right\Vert _{p,\alpha }=\left[ \mathrm{Tr}\left( \alpha
^{-1}T^{t}\alpha T\right) ^{p/2}\right] ^{1/p}.
\end{equation*}%
But
\begin{eqnarray*}
\left( \alpha ^{-1}T^{t}\alpha T\right) ^{p/2} &=&\left[ \alpha
^{-1/2}\left( \alpha ^{-1/2}T^{t}\alpha ^{1/2}\alpha ^{1/2}T\alpha
^{-1/2}\right) \alpha ^{1/2}\right] ^{p/2} \\
&=&\alpha ^{-1/2}\left[ \alpha ^{-1/2}T^{t}\alpha ^{1/2}\alpha ^{1/2}T\alpha
^{-1/2}\right] ^{p/2}\alpha ^{1/2}.
\end{eqnarray*}%
Taking trace, we obtain%
\begin{equation*}
\left\Vert T\right\Vert _{p,\alpha }=\left\Vert \alpha ^{1/2}T\alpha
^{-1/2}\right\Vert _{p},
\end{equation*}%
where $\left\Vert \cdot \right\Vert _{p}$ are the norms of operators in $%
(Z,I)$. This relation is valid also for $p=\infty .$ By using the H\"{o}lder
inequality for the Schatten norms, we obtain from (\ref{norm1})%
\begin{equation*}
\left\Vert T\right\Vert _{p,\alpha }\leq \sqrt{\left\Vert \alpha \right\Vert
\left\Vert \alpha ^{-1}\right\Vert }\left\Vert T\right\Vert _{p}\leq
2a\left\Vert T\right\Vert _{p}
\end{equation*}%
and similarly $\left\Vert T\right\Vert _{p}\leq 2a\left\Vert T\right\Vert
_{p,\alpha }.$ Hence
\begin{equation}
\frac{1}{2a}\left\Vert T\right\Vert _{p}\leq \left\Vert T\right\Vert
_{p,\alpha }\leq 2a\left\Vert T\right\Vert _{p} .  \label{palpha}
\end{equation}%
In the case of the operator $\alpha -$norm which formally corresponds to $%
p=\infty $ we have similar estimates.

We now come back to our main problem and to the quantity (\ref{ea})\ for
arbitrary states: according to (\ref{ec})
\begin{equation*}
\frac{1}{2}\mathrm{Tr\,}d(\alpha ,\beta )=\frac{1}{8}\mathrm{Tr}\left( \hat{\alpha}-\hat{\beta}\right) \left( \hat{%
\beta}^{-1}-\hat{\alpha}^{-1}\right)
\end{equation*}
\begin{equation}
\leq \frac{1}{2}\left\Vert \alpha -\beta
\right\Vert _{2}^{2}+\frac{1}{2}\left\Vert \alpha \Upsilon _{\alpha }-\beta
\Upsilon _{\beta }\right\Vert _{2}^{2}.  \label{star}
\end{equation}%
Introducing $A=2\Delta ^{-1}\alpha ,$ $B=2\Delta ^{-1}\beta $ (this time we
include factor 2 for notational convenience), the second term is evaluated
as
\begin{eqnarray*}
\frac{1}{2}\left\Vert \alpha \Upsilon _{\alpha }-\beta \Upsilon _{\beta
}\right\Vert _{2}^{2} &\leq &\frac{a^{2}}{2}\left\Vert \Upsilon _{\alpha
}-\alpha ^{-1}\beta \Upsilon _{\beta }\right\Vert _{2}^{2} \\
&\leq &2a^{4}\left\Vert \sqrt{I_{2s}+A^{-2}}-A^{-1}B\sqrt{I_{2s}+B^{-2}}%
\right\Vert _{2,\alpha }^{2},
\end{eqnarray*}%
where in the second inequality we used (\ref{palpha}). 
By (\ref{per}) both operators $\sqrt{I_{2s}+A^{-2}}$ and $A^{-1}B\sqrt{%
I_{2s}+B^{-2}}=\alpha ^{-1}\beta \sqrt{I_{2s}+B^{-2}}$ are Hermitian
positive in $Z_{\alpha }.$ Hence we can again use the inequality (\ref{21})
\begin{equation*}
\left\Vert T_{1}-T_{2}\right\Vert _{2,\alpha }^{2}\leq \left\Vert
T_{1}^{2}-T_{2}^{2}\right\Vert _{1,\alpha }
\end{equation*}%
valid for $\alpha -$positive $T_{1},T_{2},$ obtaining%
\begin{eqnarray*}
&&2a^{4}\left\Vert \sqrt{I_{2s}+A^{-2}}-A^{-1}B\sqrt{I_{2s}+B^{-2}}%
\right\Vert _{2,\alpha }^{2} \\
&\leq &2a^{4}\left\Vert I_{2s}+A^{-2}-\left( A^{-1}B\sqrt{I_{2s}+B^{-2}}%
\right) ^{2}\right\Vert _{1,\alpha } \\
&\leq &4a^{5}\left\Vert I_{2s}+A^{-2}-\left( A^{-1}B\sqrt{I_{2s}+B^{-2}}%
\right) ^{2}\right\Vert _{1}.
\end{eqnarray*}%
Expression under the norm is equal to%
\begin{equation*}
I_{2s}+A^{-2}-\left( \sqrt{I_{2s}+B^{-2}}+\left( A^{-1}B-I_{2s}\right) \sqrt{%
I_{2s}+B^{-2}}\right) ^{2}
\end{equation*}%
\begin{eqnarray*}
&=&\left( A^{-2}-B^{-2}\right) \\
&-&2\left[ \left( A^{-1}B-I_{2s}\right) \Upsilon _{\beta }\right] \circ
\Upsilon _{\beta } \\
&-&\left( \left( A^{-1}B-I_{2s}\right) \Upsilon _{\beta }\right) ^{2}.
\end{eqnarray*}

Let us evaluate trace norm of the first term of the r.h.s. By using (\ref%
{norm2}) several times, we obtain%
\begin{eqnarray*}
\left\Vert A^{-2}-B^{-2}\right\Vert _{1} &=&\left\Vert \left(
A^{-1}-B^{-1}\right) \circ \left( A^{-1}+B^{-1}\right) \right\Vert _{1} \\
&\leq &\frac{\left\Vert \alpha ^{-1}\right\Vert +\left\Vert \beta
^{-1}\right\Vert }{2}\left\Vert \left( A^{-1}-B^{-1}\right) \right\Vert _{1}
\\
&\leq &\frac{\left\Vert \alpha ^{-1}\right\Vert +\left\Vert \beta
^{-1}\right\Vert }{2}\left\Vert \beta ^{-1}\right\Vert \left\Vert
A^{-1}B-I_{2s}\right\Vert _{1} \\
&\leq &8\left( a+b\right) b\left\Vert A^{-1}B-I_{2s}\right\Vert _{1},
\end{eqnarray*}%
while by (\ref{norm2})%
\begin{equation}
\left\Vert A^{-1}B-I_{2s}\right\Vert _{1}=\left\Vert \alpha ^{-1}\beta
-I_{2s}\right\Vert _{1}\leq 4a\left\Vert \beta -\alpha \right\Vert _{1} .
\label{AB-I}
\end{equation}%
Thus%
\begin{equation}
\left\Vert A^{-2}-B^{-2}\right\Vert _{1}\leq 32(a+b)ab\left\Vert \beta
-\alpha \right\Vert _{1} .  \label{t1}
\end{equation}

For the second term we use the fact that by (\ref{aga})%
\begin{equation}
\left\Vert \Upsilon _{\beta }\right\Vert _{\beta }=\sup \frac{z^{t}\beta
\Upsilon _{\beta }z}{z^{t}\beta z}\leq 1.  \label{leq1}
\end{equation}%
Then by (\ref{palpha})
\begin{eqnarray*}
&&\left\Vert \left[ 2\left( A^{-1}B-I_{2s}\right) \Upsilon _{\beta }\right]
\circ \Upsilon _{\beta }\right\Vert _{1} \\
&\leq &2b \left\Vert \left[ 2\left( A^{-1}B-I_{2s}\right) \Upsilon _{\beta }%
\right] \circ \Upsilon _{\beta }\right\Vert _{1,\beta } \\
&\leq &4b \left\Vert A^{-1}B-I_{2s}\right\Vert _{1,\beta }.
\end{eqnarray*}%
By (\ref{AB-I}) this is evaluated as
\begin{eqnarray*}
&\leq &8b^{2}\left\Vert \alpha ^{-1}\beta -I_{2s}\right\Vert _{1} \\
&\leq &32ab^{2}\left\Vert \beta -\alpha \right\Vert _{1}.
\end{eqnarray*}

For the third term, again using (\ref{palpha}) and (\ref{leq1}),%
\begin{eqnarray*}
&&\left\Vert \left( \left( A^{-1}B-I_{2s}\right)\Upsilon _{\beta } \right)
^{2}\right\Vert _{1} \\
&\leq &2b\left\Vert \left( \left( A^{-1}B-I_{2s}\right) \Upsilon _{\beta
}\right) ^{2}\right\Vert _{1,\beta } \\
&\leq &2b\left\Vert A^{-1}B-I_{2s}\right\Vert _{1,\beta }\left\Vert
A^{-1}B-I_{2s}\right\Vert _{\beta } \\
&\leq &8b^{3}\left\Vert A^{-1}B-I_{2s}\right\Vert _{1}\left\Vert
A^{-1}B-I_{2s}\right\Vert .
\end{eqnarray*}%
From (\ref{norm2}) and (\ref{AB-I}) this is evaluated as%
\begin{eqnarray*}
&\leq &8b^{3}4a\left\Vert \beta -\alpha \right\Vert _{1}4a\left\Vert \beta
-\alpha \right\Vert \\
&=&128a^{2}b^{3}\left\Vert \beta -\alpha \right\Vert _{1}\left\Vert \beta
-\alpha \right\Vert \\
&\leq &128a^{2}b^{3}\left( a+b\right) \left\Vert \beta -\alpha \right\Vert
_{1} .
\end{eqnarray*}%
Summarizing%
\begin{eqnarray*}
\frac{1}{2}\left\Vert \alpha \Upsilon _{\alpha }-\beta \Upsilon _{\beta
}\right\Vert _{2}^{2} &\leq &4a^{5}\left[ 32(a+b)ab+32ab^{2}+128a^{2}b^{3}%
\left( a+b\right) \right] \left\Vert \beta -\alpha \right\Vert _{1} \\
&=&P(a,b)\left\Vert \beta -\alpha \right\Vert _{1}.
\end{eqnarray*}

By using (\ref{star}), (\ref{ineq6}) and (\ref{mean}) with $\alpha, \beta$
replaced by $\hat{\alpha}, \hat{\beta}$, we obtain the estimate (\ref{E3}) for the trace-norm
distance of two Gaussian states.

\section{Discussion}

\label{sec:disc}

The estimate given in \cite{lami} (theorem 2) reads in our notations
\begin{equation*}
\left\Vert \rho _{m_{1},\alpha }-\rho _{_{m_{2},\beta }}\right\Vert _{1}\leq
\sqrt{2}\left( \sqrt{N}+\sqrt{N+1}\right) \left[ \left\Vert
m_{1}-m_{2}\right\Vert _{2}+2\sqrt{\left\Vert \alpha -\beta \right\Vert _{1}}%
\right] .
\end{equation*}%
The parameter $N$ here has important physical meaning as the total mean
photon number (over all modes). Mathematically, it is expressed via the
trace norm of the covariance matrices, $N=\max \left( N_{m_{1},\alpha
},N_{_{m_{2},\beta }}\right) ,$ $N_{m,\alpha }=\frac{1}{2}\left( \mathrm{Tr}%
\left[ \alpha -\frac{1}{2}\right] +\left\Vert m\right\Vert ^{2}\right) =%
\frac{1}{2}\left( \left\Vert \alpha \right\Vert _{1}+\left\Vert m\right\Vert
^{2}-s\right) ].$ It is thoroughly adapted to the problem of learning the
energy-constrained Gaussian state. For pure states there is improved
version (lemma 4):
\begin{equation*}
\left\Vert \rho _{m_{1},\alpha }-\rho _{_{m_{2},\beta }}\right\Vert _{1}\leq
2\left( \sqrt{N+s/2}\right) \sqrt{2\left\Vert m_{1}-m_{2}\right\Vert
_{2}^{2}+2\left\Vert \alpha -\beta \right\Vert }.
\end{equation*}%
One can easily construct photon number distributions over the modes for
which $N_{m,\alpha }\,\sim \left\Vert \alpha \right\Vert _{1}\gg \left\Vert
\alpha \right\Vert =a$ and similarly for $\beta .$ In such cases our
estimates may be more stringent, notably for pure and gauge-covariant
states. Our estimates are also have the advantage of being dimension
independent, hence in principle transferrable to the case of infinite
dimensions. In that case $\alpha $ and $\beta $ are bounded operators on
one-particle Hilbert space, so the constants $a$ and $b$ entering in the
coefficients of our estimates are finite while $N$ can be easily infinite.

In the paper \cite{lami} also the lower bounds in terms of first and second
moments were derived for the trace-norm distance of two Gaussian states
(theorem 5). It would be interesting to investigate whether the second
inequality in (\ref{second}) could be used for a similar purpose.

\section{Appendix: the case of fermionic Gaussian states of CAR}

Here we briefly recall some results from our paper \cite{tmf2} where
the states overlap was computed for general fermionic Gaussian states
of canonical anticommutation relations (CAR).

Let $H$ be a real even- or infinite-dimensional Hilbert space with inner product $(f,g)$.
Let $\mathfrak{A}(H)$ the C$^{\ast }$-algebra of CAR generated by the elements $B(f),\,f\in H,$ where $%
f\rightarrow B(f)$ is a linear map and
\begin{equation*}
B(f)B(g)+B(g)B(f)=2(f,g)I.
\end{equation*}%
A state $\rho \left( \cdot \right) $ on $\mathfrak{A}(H)$ is called
quasifree (also fermionic Gaussian) if it is even and
\begin{equation}
\rho \left( B(f_{1})\dots B(f_{2n})\right) =\sum \left( -1\right) ^{p}\rho
\left( B(f_{i_{1}})B(f_{j_{1}})\right) \dots \rho \left(
B(f_{i_{n}})B(f_{j_{n}})\right) ,  \label{even}
\end{equation}%
where the sum is over all permutations $(1,2,\dots ,2n)\rightarrow
(i_{1},j_{1},\dots ,i_{n},j_{n})$ $\ $such that$\ \ i_{k}<j_{k},$ $\ \
i_{1}<i_{2}<\ \dots <i_{n}.$ Quasifree state defines an operator $A$ on $H$
by the formula
\begin{equation*}
\rho \left( B(f)B(g)\right) =(f,g)+i(Af,g).
\end{equation*}%
The operator  $A$ satisfies

\begin{equation}
A^{\ast }=-A,\ \left\Vert A\right\Vert \leq 1,  \label{A}
\end{equation}%
and it is called covariance operator of the the state $\rho \left( \cdot
\right) .$Conversely, any such operator defines a state according to (\ref{A}%
), (\ref{even}) a quasifree state which is denoted $\rho _{A}\left( \cdot
\right) .$ We denote by $\rho _{A}$ the density operator of the state $\rho
_{A}(\cdot ).$

One has $0\leq A^{\ast }A=-A^{2}\leq I.$ The operator $I+A^{2}$ is
symmetric, $0\leq I+A^{2}\leq I,$ hence $\sqrt{I+A^{2}}$ is defined and $%
0\leq \sqrt{I+A^{2}}\leq I.$ The state $\rho _{A}$ is pure iff $I+A^{2}=0.$
For the proof of the following results see \cite{tmf2}.

\begin{lemma}
\emph{Let $A,B$ be operators in $H$ satisfying (\ref{A}), then the states overlap of $\rho _{A}, \rho _{B}$ is}
\begin{equation}
\mathrm{Tr}\sqrt{\rho _{A}}\sqrt{\rho _{B}}=\sqrt[4]{\det S^{\ast }S}
\label{over}
\end{equation}%
\emph{where}
\begin{eqnarray*}
S&=&\frac{1}{2}[ \left( I+\sqrt{I+B^{2}}\right) ^{1/2}\left( I+\sqrt{I+A^{2}}%
\right) ^{1/2} \\
&-&B\left( I+\sqrt{I+B^{2}}\right) ^{-1/2}A\left( I+\sqrt{I+A^{2}}\right)
^{-1/2}]
\end{eqnarray*}
\end{lemma}

\begin{lemma}
$0\leq S^{\ast }S\leq I.$
\end{lemma}

Introducing $s(A,B)=\mathrm{Tr}\left( I-S^{\ast }S\right) $ we have the
following estimates for the states overlap:

\begin{equation}
\sqrt[4]{1-s(A,B)}\leq \mathrm{Tr}\sqrt{\rho _{A}}\sqrt{\rho _{B}}\leq \exp
\left( -\frac{s(A,B)}{4}\right) ,  \label{fover}
\end{equation}%
provided $s(A,B)\leq 1.$ We have
\begin{eqnarray*}
s(A,B) &=&\frac{1}{4}\mathrm{Tr}\left[ \left( A-B\right) ^{\ast }\left(
A-B\right) +\left( \sqrt{I+A^{2}}-\sqrt{I+B^{2}}\right) ^{2}\right]  \\
&=&\frac{1}{4}\left[ \left\Vert A-B\right\Vert _{2}^{2}+\left\Vert \sqrt{%
I+A^{2}}-\sqrt{I+B^{2}}\right\Vert _{2}^{2}\right].  \end{eqnarray*}
The second term here can be evaluated similarly to the second term in (\ref{sqrt})
leading to
\begin{eqnarray*}
s(A,B)&\leq &\frac{1}{4}\left[ \left\Vert A-B\right\Vert _{2}^{2}+\left\Vert
A^{2}-B^{2}\right\Vert _{1}\right]  \\
&\leq &\frac{1}{4}\left[ \left\Vert A-B\right\Vert _{2}^{2}+\left(
\left\Vert A\right\Vert +\left\Vert B\right\Vert \right) \left\Vert
A-B\right\Vert _{1}\right]  \\
&=&\frac{1}{4}\left[ \left\Vert A-B\right\Vert _{2}^{2}+2\left\Vert
A-B\right\Vert _{1}\right] .
\end{eqnarray*}
From the first inequality in (\ref{fover}) we then obtain
\begin{eqnarray}\notag
\left[\mathrm{Tr}\sqrt{\rho _{A}}\sqrt{\rho _{B}}\right]^2&\geq&
\sqrt{1-s(A,B)}\sim 1-\frac{1}{2}s(A,B)\\&\geq& 1-\frac{1}{8}\left[ \left\Vert A-B\right\Vert _{2}^{2}+2\left\Vert
A-B\right\Vert _{1}\right] .\label{under}
\end{eqnarray}

When both $\rho _{A}$ and $\rho _{B}$ are pure, (\ref{over}) reduces to%
\begin{eqnarray*}
\mathrm{Tr}\sqrt{\rho _{A}}\sqrt{\rho _{B}} &=&\sqrt[4]{\det \left( \frac{%
I-AB}{2}\right) \left( \frac{I-BA}{2}\right) } \\
&=&\sqrt[4]{\det \left[ -\left( \frac{A+B}{2}\right) ^{2}\right] } \\
&=&\sqrt{\det \left( \frac{A+B}{2}\right) },
\end{eqnarray*}%
also $s(A,B)=\frac{1}{4}\left\Vert A-B\right\Vert _{2}^{2}.$ By using the
estimate (\ref{basic3}) we get%
\begin{eqnarray*}
\left\Vert \rho _{A}-\rho _{B}\right\Vert _{1} &=&2\sqrt{1-\mathrm{Tr}\sqrt{%
\rho _{A}}\sqrt{\rho _{B}}} \\
&=&2\sqrt{1-\sqrt{\det \left( \frac{A+B}{2}\right) }} \\
&\leq &2\sqrt{1-\sqrt[4]{1-\frac{1}{4}\left\Vert A-B\right\Vert _{2}^{2}}}
\end{eqnarray*}%
provided $\left\Vert A-B\right\Vert _{2}\leq 2.$ For small $\left\Vert
A-B\right\Vert _{2}$ this is $\sim \frac{1}{2}\left\Vert A-B\right\Vert _{2}.
$ The second inequality in (\ref{fover}) implies a lower bound:%
\begin{equation*}
\left\Vert \rho _{A}-\rho _{B}\right\Vert _{1}\geq 2\left( 1-\exp \left( -%
\frac{\left\Vert A-B\right\Vert _{2}^{2}}{16}\right) \right) .
\end{equation*}

For further results concerning the trace-norm distance of fermionic Gaussian states
cf. recent paper \cite{eis}.

\textbf{Acknowledgment.} The author is grateful to M.E. Shirokov for comments
helping to improve the presentation.


\end{document}